\documentclass[aps,prb,superscriptaddress,nofootinbib,longbibliography,twocolumn]{revtex4-2}
\usepackage{bm}
\usepackage{lmodern}

\usepackage[utf8]{inputenc}
\usepackage{amsmath}
\usepackage{amssymb}
\usepackage{upgreek}
\usepackage{graphicx}
\usepackage{epstopdf}
\usepackage[dvipsnames]{xcolor}
\usepackage{enumerate}
\usepackage{gensymb}
\usepackage{ulem}
\usepackage[T1]{fontenc}
\newcommand{\stkout}[1]{\ifmmode\text{\sout{\ensuremath{#1}}}\else\sout{#1}\fi}
\usepackage[english]{babel}
\usepackage{braket}
\usepackage{natbib}
\usepackage{float}
\usepackage{stmaryrd}
\usepackage{mathtools}

\usepackage[misc]{ifsym}

\newcommand\blankfootnote[1]{%
  \let\thefootnote\relax\footnotetext{#1}%
  \let\thefootnote\svthefootnote%
}

\usepackage{bibunits}

\usepackage{hyperref}
\usepackage{lipsum}
\begin{document}

%\title{Coherent shuttling of a hole spin qubit in a 2D semiconductor quantum dotarray}
% \title{Coherent spin qubit shuttling through quantum dots in germanium}
% \title{Coherent shuttling of a hole spin qubit through multiple quantum dots}
%\title{\CD{Coherent shuttling of a hole spin qubit through quantum dots}}
\title{Coherent spin qubit shuttling through germanium quantum dots} %% This is still something to think about, include holes or germanium or not?

\author{Floor van Riggelen-Doelman}
\affiliation{QuTech and Kavli Institute of Nanoscience, Delft University of Technology, PO Box 5046, 2600 GA Delft, The Netherlands}
\author{Chien-An Wang}
\affiliation{QuTech and Kavli Institute of Nanoscience, Delft University of Technology, PO Box 5046, 2600 GA Delft, The Netherlands}
\author{Sander L. de Snoo}
\affiliation{QuTech and Kavli Institute of Nanoscience, Delft University of Technology, PO Box 5046, 2600 GA Delft, The Netherlands}
\author{William I. L. Lawrie}
\affiliation{QuTech and Kavli Institute of Nanoscience, Delft University of Technology, PO Box 5046, 2600 GA Delft, The Netherlands}
\author{Nico W. Hendrickx}
\affiliation{QuTech and Kavli Institute of Nanoscience, Delft University of Technology, PO Box 5046, 2600 GA Delft, The Netherlands}
\author{Maximilian Rimbach-Russ}
\affiliation{QuTech and Kavli Institute of Nanoscience, Delft University of Technology, PO Box 5046, 2600 GA Delft, The Netherlands}
\author{Amir Sammak}
\affiliation{QuTech and Netherlands Organisation for Applied Scientific Research (TNO), Delft, The Netherlands}
\author{Giordano Scappucci}
\affiliation{QuTech and Kavli Institute of Nanoscience, Delft University of Technology, PO Box 5046, 2600 GA Delft, The Netherlands}
\author{Corentin Déprez}
\affiliation{QuTech and Kavli Institute of Nanoscience, Delft University of Technology, PO Box 5046, 2600 GA Delft, The Netherlands}
\author{Menno Veldhorst}
\affiliation{QuTech and Kavli Institute of Nanoscience, Delft University of Technology, PO Box 5046, 2600 GA Delft, The Netherlands}

\date{\today}

\begin{abstract}
%A quantum bus has promise to interconnect qubit registers for networked quantum computing. Semiconductor quantum dot qubits have seen significant progress in the high-fidelity operation of small qubit registers but establishing a compelling quantum link remains a challenge. Here, we show that a spin qubit can be shuttled through multiple quantum dots while preserving its quantum information. Remarkably, we achieve these results using hole spin qubits in germanium, despite the presence of strong spin-orbit interaction. We accomplish the shuttling of spin basis states over effective lengths beyond 300$~\upmu$m and demonstrate the coherent shuttling of superposition states over effective lengths corresponding to 9$~\upmu$m, which we can extend to 49$~\upmu$m by incorporating dynamical decoupling. These findings indicate qubit shuttling as an effective approach to route qubits within registers and to establish quantum links between registers.

Quantum links can interconnect qubit registers and are therefore essential in networked quantum computing. Semiconductor quantum dot qubits have seen significant progress in the high-fidelity operation of small qubit registers but establishing a compelling quantum link remains a challenge. Here, we show that a spin qubit can be shuttled through multiple quantum dots while preserving its quantum information. Remarkably, we achieve these results using hole spin qubits in germanium, despite the presence of strong spin-orbit interaction. We accomplish the shuttling of spin basis states over effective lengths beyond 300$~\upmu$m and demonstrate the coherent shuttling of superposition states over effective lengths corresponding to 9$~\upmu$m, which we can extend to 49$~\upmu$m by incorporating dynamical decoupling. These findings indicate qubit shuttling as an effective
approach to route qubits within registers and to establish quantum links between registers.
% 140 words

\end{abstract}

% Count Word 27/07/2023 (No abstract, no method, no captions)
% 2771
% Headers:
% 8
% Math Inline:
% 110
% Math Display:
% 0

\maketitle

\maketitle

\section*{Introduction}

The envisioned approach for semiconductor spin qubits towards fault-tolerant quantum computation centers on the concept of quantum networks, where qubit registers are interconnected via quantum links~\cite{vandersypen2017}. Significant progress has been made in controlling few-qubit registers~\cite{Hendrickx2021, Philips2022}. Recent efforts have led to demonstrations of high fidelity single- and two-qubit gates~\cite{Lawrie2023, Mills2022}, quantum logic above one Kelvin~\cite{Petit2020,Yang2020,Camenzind2022} and operation of a 16 quantum dot array~\cite{Borsoi2022}. However, scaling up to larger qubit numbers requires changes in the device architecture~\cite{vanMeter2013,Wecker2014,Reiher2017, Franke2019}. 

Inclusion of short-range and mid-range quantum links could be particularly effective to establish scalability, addressability, and qubit connectivity. The coherent shuttling of electron or hole spins is an appealing concept for the integration of such quantum links in spin qubit devices. Short-range coupling, implemented by shuttling a spin qubit through quantum dots in an array, can provide flexible qubit routing and local addressability~\cite{Li2018, Noiri2022}. Moreover, it allows to increase connectivity beyond nearest-neighbour coupling and decrease the number of gates needed to execute algorithms. Mid-range links, implemented by shuttling spins through a multitude of quantum dots, may entangle distant qubit registers for networked computing and allow for qubit operations at dedicated locations~\cite{Taylor2005,Li2018,Boter2022,Kunne2023}. Furthermore, such quantum buses could provide space for the integration of on-chip control electronics~\cite{vandersypen2017}, depending on their footprint. 

The potential of shuttling-based quantum buses has stimulated research on shuttling electron charge~\cite{Mills2019,Seider2022,Xue2023} and spin~\cite{Flentje2017,Fujita2017,Mortemousque2021,Mortemousque2021b,Jadot2021,Yoneda2021,Noiri2022,Zwerver2023,Struck2023}. While nuclear spin noise prevents high-fidelity qubit operation in gallium arsenide, demonstrations of coherent transfer of individual electron spins through quantum dots are encouraging ~\cite{Flentje2017,Fujita2017,Mortemousque2021,Mortemousque2021b,Jadot2021}. In silicon, qubits can be operated with high-fidelity and this has been employed to displace a spin qubit in a double quantum dot~\cite{Yoneda2021, Noiri2022}. Networked quantum computers, however, will require integration of qubit control and shuttling through quantum dots.

Meanwhile, quantum dots defined in strained germanium (Ge/SiGe) heterostructures have emerged as a promising platform for hole spin qubits~\cite{Sammak2019,Scappucci2020}. The high quality of the platform allowed for rapid development of single spin qubits~\cite{Hendrickx2020a,Hendrickx2020b}, singlet-triplet qubits~\cite{Jirovec2021,Jirovec2022,Wang2023_RVB}, a four qubit processor~\cite{Hendrickx2021}, and a 4$\times$4 quantum dot array with shared gate control~\cite{Borsoi2022}. While the strong spin orbit interaction allows for fast and all-electrical control, the resulting anisotropic $g$-tensor~\cite{Scappucci2020, Hendrickx2023} complicates the spin dynamics and may challenge the feasibility of a quantum bus.

Here, we demonstrate that spin qubits can be shuttled through quantum dots. These experiments are performed with two hole spin qubits in a 2$\times$2 germanium quantum dot array. Importantly, we operate in a regime where we can implement single qubit logic and coherently transfer spin qubits to adjacent quantum dots. Furthermore, by performing experiments with precise voltage pulses and sub-nanosecond time resolution, we can mitigate finite qubit rotations induced by spin-orbit interactions. In these optimized sequences we find that the shuttling performance is limited by dephasing and can be extended through dynamical decoupling. 

\begin{figure*}
\centering
\includegraphics[width=1\textwidth]{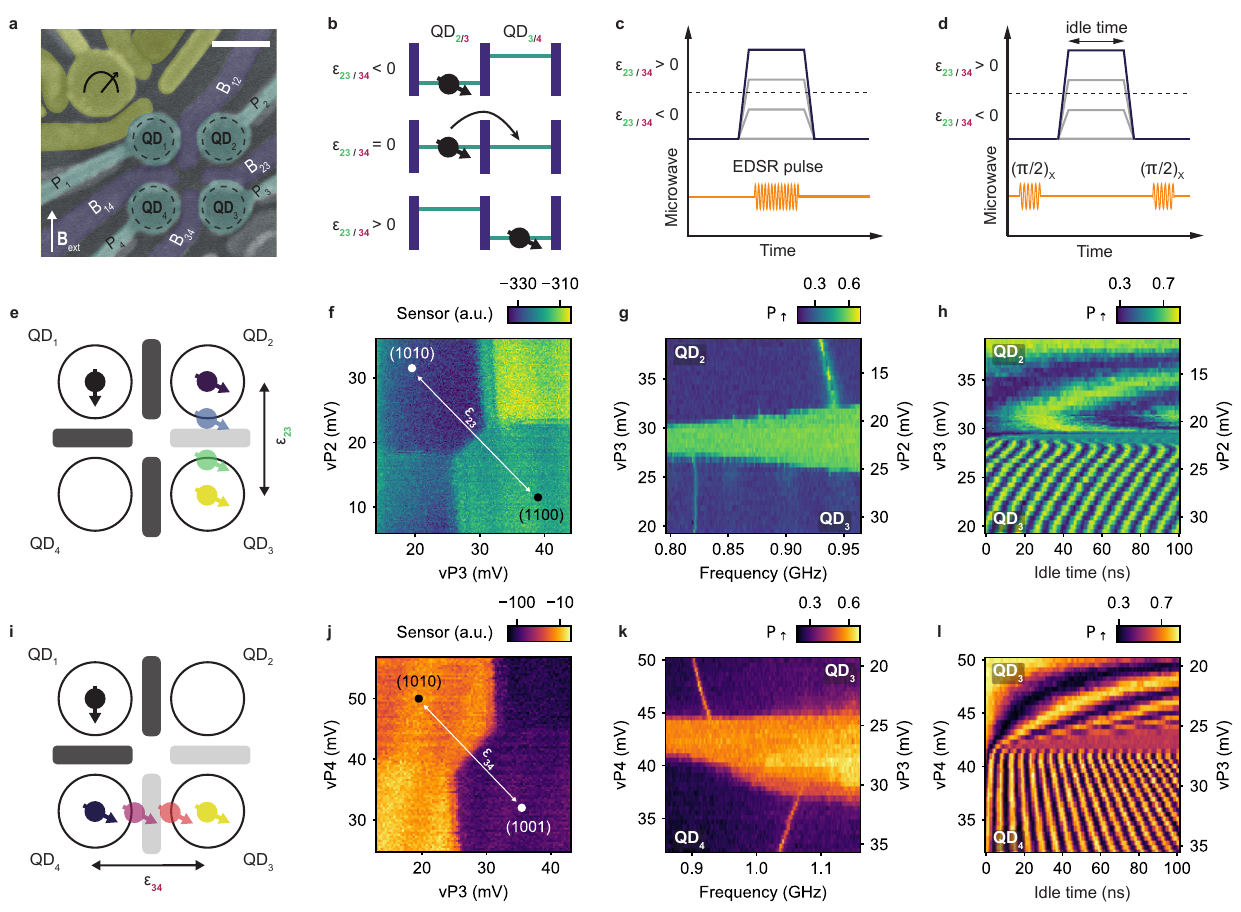}
\caption{\textbf{Coherent shuttling of hole spin qubits in germanium double quantum dots.} \textbf{a}, A false colored scanning electron microscope image of a similar device to the one used in this work. The quantum dots are formed under the plunger gates (light blue) and separated by barrier gates (dark blue) which control the tunnel couplings. A single hole transistor is defined by the yellow gates and is used as charge sensor. The scale bar corresponds to 100~nm. \textbf{b}, Schematic showing the principle of bucket brigade mode shuttling. The detuning energy $\epsilon_{23/34}$ between the two quantum dots is progressively changed such that it becomes energetically favorable for the hole to tunnel from one quantum dot to another. \textbf{c}, Schematic of the pulses used for the shuttling experiments shown in (g) and (k), where the resonance frequency of the qubit is probed after the application of a detuning pulse using a 4$~\upmu$s EDSR pulse. \textbf{d}, Schematic of the pulses used for coherent shuttling experiments of which the results are shown in (h) and (l). The qubit is prepared in a superposition state using a $\pi/2$ pulse and is transferred to the empty quantum dot with a detuning pulse of varying amplitude, and then brought back to its initial position after an idle time. After applying another $\pi/2$ pulse we readout the spin state. \textbf{e}, \textbf{i}, Schematic illustrating the shuttling of a spin qubit between QD$_{2}$ and QD$_{3}$ (e) and between QD$_{3}$ and QD$_{4}$ (i). \textbf{f}, \textbf{j}, Charge stability diagrams of QD$_{2}$-QD$_{3}$ (f) and  QD$_{3}$-QD$_{4}$ (j). To shuttle the qubit from one site to another, the virtual plunger gate voltages are varied along the detuning axis (white arrow), which crosses the interdot charge transition line. \textbf{g}, \textbf{k}, Probing of the resonance frequency along the detuning axis for the double quantum dot QD$_{2}$-QD$_{3}$ (g) and QD$_{3}$-QD$_{4}$ (k). The resonance frequencies of the spin in the different quantum dots are clearly visible, indicating the possibility to shuttle a hole while preserving its spin polarization. Nearby the charge transition, the resonance frequency cannot be resolved due to a combination of effects discussed in Supplementary Note 1. \textbf{h}, \textbf{l}, Coherent free evolution of a qubit during the shuttling between QD$_{2}$-QD$_{3}$ (h) and QD$_{3}$-QD$_{4}$ (l). Since the Larmor frequency varies along the detuning axes, the qubit initialized in a superposition state acquires a phase that varies with the idle time resulting in oscillations in the spin-up P$_{\uparrow}$ probabilities.}
\label{fig:Fig1}
\end{figure*}

\section*{Coherent shuttling of single hole spin qubits}

Fig.~\ref{fig:Fig1}.a shows a germanium 2$\times$2 quantum dot array identical to the one used in the experiment~\cite{Hendrickx2021}. The chemical potentials and the tunnel couplings of the quantum dots are controlled with virtual gates (vP$_{\rm i}$, vB$_{\rm ij}$), which consist of combinations of voltages on the plunger gates and the barrier gates. We operate the device with two spin qubits in quantum dots QD$_{1}$ and QD$_{2}$ and initialised the $\ket{\downarrow\downarrow}$ state (see Methods). We use the qubit in QD$_{1}$ as an ancilla to readout the hole spin in QD$_{2}$, using latched Pauli spin blockade \cite{Studenikin2012,HarveyCollard2018,Hendrickx2021}. The other qubit starts in QD$_{2}$ and is shuttled to the other quantum dots by changing the detuning energies ($\epsilon_{23/34}$) between the quantum dots (Fig.~\ref{fig:Fig1}.b, e and i). 
The detuning energies are varied by pulsing the plunger gate voltages as illustrated in Fig.~\ref{fig:Fig1}.f and j. Additionally, we increase the tunnel couplings between QD$_{2}$-QD$_{3}$ and QD$_{3}$-QD$_{4}$ before shuttling to allow for adiabatic charge transfer.

\begin{figure*}
\centering
\includegraphics[width=1\textwidth]{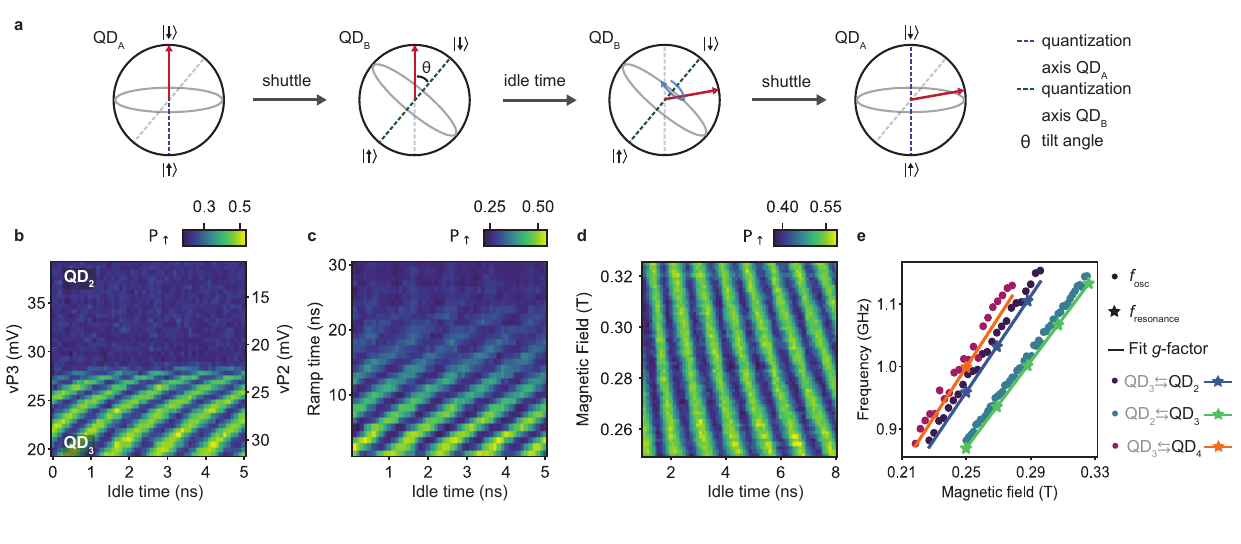}
\caption{\textbf{Rotations induced while shuttling by the difference in quantization axes.} \textbf{a}, Schematic explaining the effect of the change in quantization axis direction that the qubit experiences during the shuttling process. The difference in quantization axis between quantum dots is caused by the strong spin-orbit interaction. \textbf{b}, Oscillations induced by the change in quantization axis while shuttling diabatically a qubit in a $\ket{\downarrow}$ state between QD$_2$ and QD$_3$. Ramp times of 4 ns are used for the detuning pulses. \textbf{c}, Oscillations due to the change in quantization axis at a fixed point in detuning, as function of the voltage pulse ramp time used to shuttle the spin. When the ramp time is long enough, typically above 30 ns, the spin is shuttled adiabatically and the oscillations vanish. \textbf{d}, Magnetic-field dependence of the oscillations induced by the difference in quantization axis. \textbf{e}, Frequency of the oscillations $f_{\rm osc}$ induced by the change in quantization axis as a function of magnetic field for different shuttling processes. The oscillation frequency $f_{\rm osc}$ for QD$_3$ is extracted from measurements displayed in (d) (and similar experiments for the other quantum dot pairs) and is plotted with points. $f_{\rm osc}$ scales linearly with the magnetic field. Comparing $f_{\rm osc}$ with resonance frequencies measured using EDSR pulses (data points depicted with stars) reveals that $f_{\rm osc}$ is given by the Larmor frequency of the quantum dot towards which the qubit is shuttled (black label).}
\label{fig:Fig2}

\end{figure*}

The $g$-tensor of hole spin qubits in germanium is sensitive to the local electric field. Therefore, the Larmor frequency ($f_{\rm{L}}$) is different in each quantum dot~\cite{Hendrickx2020a,Hendrickx2020b,Jirovec2021}. We exploit this effect to confirm the shuttling of a hole spin from one quantum dot to another. In Fig.~\ref{fig:Fig1}.c. we show the experimental sequence used to measure the qubit resonance frequency, while changing the detuning to transfer the qubit. Fig~\ref{fig:Fig1}.g (k) shows the experimental results for spin transfers from QD$_2$ to QD$_3$ (QD$_3$ to QD$_4$). Two regions can be clearly distinguished in between which $f_{\rm L}$ varies by 110 (130) MHz. This obvious change in $f_{\rm L}$ clearly shows that the hole is shuttled from QD$_2$ to QD$_3$ (QD$_3$ to QD$_4$) when applying a sufficiently large detuning pulse. To investigate whether such transfer is coherent, we probe the free evolution of qubits prepared in a superposition state after applying a detuning pulse (Fig.~\ref{fig:Fig1}.d)~\cite{Yoneda2021}. The resulting coherent oscillations are shown in Fig.~\ref{fig:Fig1}.h (l). They are visible over the full range of voltages spanned by the experiment and arise from a phase accumulation during the idle time. Their frequency $f_{\rm osc}$ is determined by the difference in resonance frequency between the starting and end point in detuning as shown in Supplementary Figure~1. The abrupt change in $f_{\rm osc}$ marks the point where the voltage pulse is sufficiently large to transfer the qubit from QD$_2$ to QD$_3$ (QD$_3$ to QD$_4$). These results clearly demonstrate that single hole spin qubits can be coherently transferred.

\section*{The effect of strong spin-orbit interaction on spin shuttling}

The strong spin-orbit interaction in our system has a significant impact on the spin dynamics during the shuttling. It appears when shuttling a qubit in a $\ket{\downarrow}$ state between QD$_2$ and QD$_3$ using fast detuning pulses with voltage ramps of 4~ns. Doing this generates coherent oscillations shown in Fig.~\ref{fig:Fig2}.b that appear only when the qubit is in QD$_3$. They result from the strong spin-orbit interaction and the use of an almost in-plane magnetic field~\cite{LawriePhD}. In this configuration, the direction of the spin quantization axis depends strongly on the local electric field~\cite{Mutter2021,Bosco2021,Jirovec2022,wang2022_modelling,Hendrickx2023} and can change significantly between neighbouring quantum dots. Therefore, a qubit in a spin basis state in QD$_2$ becomes a superposition state in QD$_3$ when diabatically shuttled. Consequently, the spin precesses around the quantization axis of QD$_3$ until it is shuttled back (Fig.~\ref{fig:Fig2}.a). This leads to qubit rotations and the aforementioned oscillations. 

While these oscillations are clearly visible for voltage pulses with ramp times $t_{\rm ramp}$ of few nanoseconds, they fade as the ramp times are increased, as shown in Fig.~\ref{fig:Fig2}.c, and vanish for $t_{\rm ramp}>30$~ns. The qubit is transferred adiabatically  and can follow the change in quantization axis and therefore remains in the spin basis state in both quantum dots.
%The qubit is then transferred adiabatically with respect to the change in quantization axis and remains in the spin basis state in both quantum dots.  
From the visibility of the oscillations, we estimate that the quantization axis of QD$_3$ (QD$_4$) is tilted by at least 42$\degree$ (33$\degree$) compared to the quantization axis of QD$_2$ (QD$_3$). These values are corroborated by independent estimations made by fitting the evolution of $f_{\rm L}$ along the detuning axes (see Supplementary Note 2).

Fig.~\ref{fig:Fig2}.d and Fig.~\ref{fig:Fig2}.e display the magnetic field dependence of the oscillations generated by diabatic shuttling. Their frequencies $f_{\rm osc}$ increase linearly with the field and match the Larmor frequencies $f_{\rm L}$ measured for a spin in the target quantum dot. This is consistent with the explanation that the oscillations are due to the spin precessing around the quantization axis of the second quantum dot. 

\begin{figure*}
\centering
\includegraphics[width=1\textwidth]{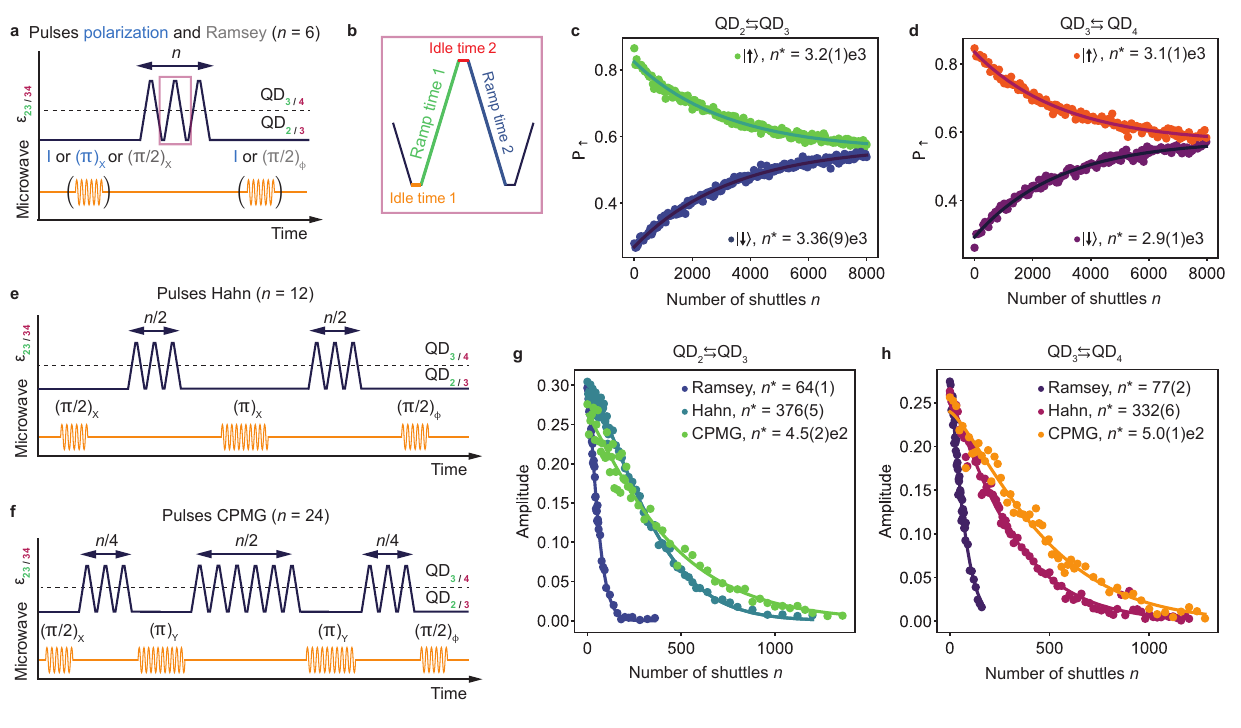}
\caption{\textbf{Quantifying the performance for the shuttling in double quantum dots.} \textbf{a}, Schematic of the pulse sequence used for quantifying the performance of shuttling basis states (blue) or a superposition state (grey). The spin qubit is prepared in the quantum dot where the shuttling experiment starts, by either applying an identity gate (shuttling a $\ket{\downarrow}$ state), a $(\pi)_\text{X}$ pulse (shuttling a $\ket{\uparrow}$ state) or ($\pi$/2)$_\text{X}$ pulse (shuttling a superposition state, also referred to as Ramsey shuttling experiments). Detuning pulses are applied to the plunger gates to shuttle the hole from one quantum dot to another, back and forth, and finally the appropriate pulses are applied to prepare for readout. Moving the qubit from one quantum dot to another is counted as one shuttling event $n=1$. Since the hole always needs to be shuttled back for readout, $n$ is always an even number. The schematic shows an example for $n=6$. \textbf{b}, Zoom-in on the detuning pulses used for the shuttling. To make an integer number of $2\pi$ rotation(s) around the quantization axis of the second quantum dot, all ramp and idle times in the pulse need to be optimized. \textbf{c, d}, Spin-up probabilities P$_{\uparrow}$ measured after shuttling $n$ times a qubit prepared in a spin basis state between QD$_2$ and QD$_3$ (c) and between QD$_3$ and QD$_4$ (d). The decay of P$_{\uparrow}$ as a function of $n$ is fitted to an exponential function P$_{\uparrow}=\text{P}_0\exp(-n/n^{\ast})+\text{P}_{\rm sat}$. \textbf{e}, Pulse sequence used for implementing a Hahn echo shuttling experiment. In the middle of the shuttling experiment, an echo pulse $(\pi)_\text{X}$ is applied in the quantum dot where the spin qubit was initially prepared. Example for $n=12$. \textbf{f}, Pulse sequence for a CPMG shuttling experiment. Two $(\pi)_\text{Y}$ pulses are inserted between the shuttling pulses. Example for $n=24$. \textbf{g, h}, Performance of the shuttling of superposition state between QD$_2$ and QD$_3$ (g) and QD$_2$ and QD$_3$ (h) for different shuttling sequences. The decay of the coherent amplitude $A$ of the superposition state are fitted by $A_0\exp{(-({n/n^{\ast}})^\alpha)}$ where $\alpha$ is a fitting parameter.}
\label{fig:Fig3}
\end{figure*}

\section*{Shuttling performance}

To quantify the performance of shuttling a spin qubit, we implement the experiments depicted in Fig.~\ref{fig:Fig3}.a, e and f ~\cite{Yoneda2021,Noiri2022} and study how the state of a qubit evolves depending on the number of subsequent shuttling events. For hole spins in germanium, it is important to account for rotations induced by the spin-orbit interaction. This can be done by aiming to avoid unintended rotations, or by developing methods to correct them. An example of the first approach is transferring the spin qubits adiabatically. This implies using voltage pulses with ramps of tenths of nanoseconds, which are significant with respect to the dephasing time. However, this strongly limits the shuttling performance (see Supplementary Figure~5). Instead, we can mitigate rotations by carefully tuning the duration of the voltage pulses, such that the qubit performs an integer number of $2\pi$ rotations around the quantization axis of the respective quantum dot. This approach is demanding, as it involves careful optimization of the idle times in each quantum dot as well as the ramp times, as depicted in Fig.~\ref{fig:Fig3}.b. However, it allows for fast shuttling, with ramp times of typically 4~ns and idle times of 1~ns, significantly reducing the dephasing experienced by the qubit during the shuttling. We employ this strategy in the rest of our experiments.

We first characterize the fidelity of shuttling spin basis states. % by transferring multiple times qubits prepared in $\ket{\uparrow}$ or $\ket{\downarrow}$ states between the quantum dots. 
We do this by preparing a qubit in a $\ket{\uparrow}$ or $\ket{\downarrow}$ state and transferring it multiple times between the quantum dots. Fig.~\ref{fig:Fig3}.c and d display the spin-up fraction $\rm P_{\uparrow}$ measured as a function of the number of shuttling steps $n$. The probability of ending up in the initial state shows a clear exponential dependence on $n$. No oscillations of $\rm P_{\uparrow}$ with $n$ are visible, confirming that the pulses have been successfully optimized to account for unwanted spin rotations. We find for the shuttling of basis states characteristic decay constants $n^{\ast}=$ 3000 shuttlings, corresponding to polarization transfer fidelities $F=\exp(-1/n^{\ast})\simeq 99.97~\%$. This is similar to the fidelities reached in silicon devices~\cite{Yoneda2021,Noiri2022}, despite the anisotropic $g$-tensors due to the strong spin-orbit interaction in our platform. 

We now focus on the performance of coherent shuttling. We prepare a superposition state via an EDSR $(\pi/2)_{\rm X}$ pulse, shuttle the qubit, apply another $\pi/2$ pulse and measure the spin state. Importantly, one must account for $\hat{z}$-rotations experienced by the qubits during the experiments. Therefore, we vary the phase of the EDSR pulse $\upphi$ for the second $\pi/2$ pulse. For each $n$, we then extract the amplitude $A$ of the $\rm P_{\uparrow}$ oscillations that appear as function of $\upphi$~\cite{Yoneda2021,Noiri2022}. Fig.~\ref{fig:Fig3}.g, h show the evolution of $A$ as a function of $n$ for shuttling between adjacent quantum dots. We fit the experimental results using $A_0\exp{(-({n/n^{\ast}})^\alpha)}$ and find characteristic decay constants $n^{\ast}_{23}=64 \pm1 $ and $n^{\ast}_{34}=77 \pm 2$. Remarkably, these numbers compare favourably to $n^{\ast}\simeq50$ measured in a SiMOS electron double quantum dot~\cite{Yoneda2021}, where the spin-orbit coupling is weak.

% The exponents, $\alpha_{23} = 1.36 \pm 0.05$ and $\alpha_{34} = 1.28 \pm 0.06$, reveal that the decays are not exponential. This means that the error is not the same for every $n$ and indicates that apart from the errors induced by the shuttling itself, there are other mechanisms that induce errors. 

% We suspect that dephasing during free evolution plays an important role. We investigate this by including a Hahn echo in the shuttling sequence, with the echo pulse $(\pi)_{\rm X}$ applied in the middle of the sequence as displayed in Fig.~\ref{fig:Fig3}.e. Fig.~\ref{fig:Fig3}.g and h show the experimental results and it is clear that the performance is improved significantly. We can extend the shuttling by a factor four to five, reaching a characteristic decay of more than 300 shuttlings. Similarly, the use of CMPG sequences incorporating two decoupling $(\pi)_{\rm Y}$ pulses (Fig.~\ref{fig:Fig3}.f) allows further, though modest, improvements. These enhancements in the shuttling performance confirm that dephasing is limiting the shuttling performance.

% Both the non-exponential decay and the improvement due to spin echo are in contrast to observations in SiMOS~\cite{Yoneda2021}. We speculate that the origin of the difference is two-fold. Firstly, due to the strong spin-orbit interaction, the spin is more sensitive to electrical noise, which results in a shorter dephasing time. Secondly, the excellent control over the potential landscape in germanium allows to minimize errors due to the shuttling itself.

The exponents, $\alpha_{23} = 1.36 \pm 0.05$ and $\alpha_{34} = 1.28 \pm 0.06$, reveal that the decays are not exponential. This contrasts with observations in silicon~\cite{Yoneda2021,Noiri2022}, and suggests that the shuttling of hole spins in germanium is limited by other mechanisms. Two types of errors can be distinguished: those induced by the shuttling processes and errors due to the dephasing during free evolution. To investigate the effect of the latter, we  modify the shuttling sequence and include a $(\pi)_{\rm X}$ echoing pulse in the middle as displayed in Fig.~\ref{fig:Fig3}.e. Fig.~\ref{fig:Fig3}.g and h show the experimental results and it is clear that in germanium the coherent shuttling performance is improved significantly using an echo pulse: we can extend the shuttling by a factor of four to five, reaching a characteristic decay of more than 300 shuttles. Similarly, the use of CPMG sequences incorporating two decoupling $(\pi)_{\rm Y}$ pulses (Fig.~\ref{fig:Fig3}.f) allows further, though modest, improvements. These enhancements in the shuttling performance confirm that dephasing is limiting the shuttling performance contrary to observations in SiMOS~\cite{Yoneda2021}. We speculate that the origin of the difference is two-fold. Firstly, due to the stronger spin-orbit interaction, the spin is more sensitive to charge noise, resulting in a shorter dephasing times~\cite{Stano2022}. Secondly, the excellent control over the potential landscape in germanium allows minimizing the errors which are due to the shuttling itself.

\begin{figure*}
\centering
\includegraphics[width=\textwidth]{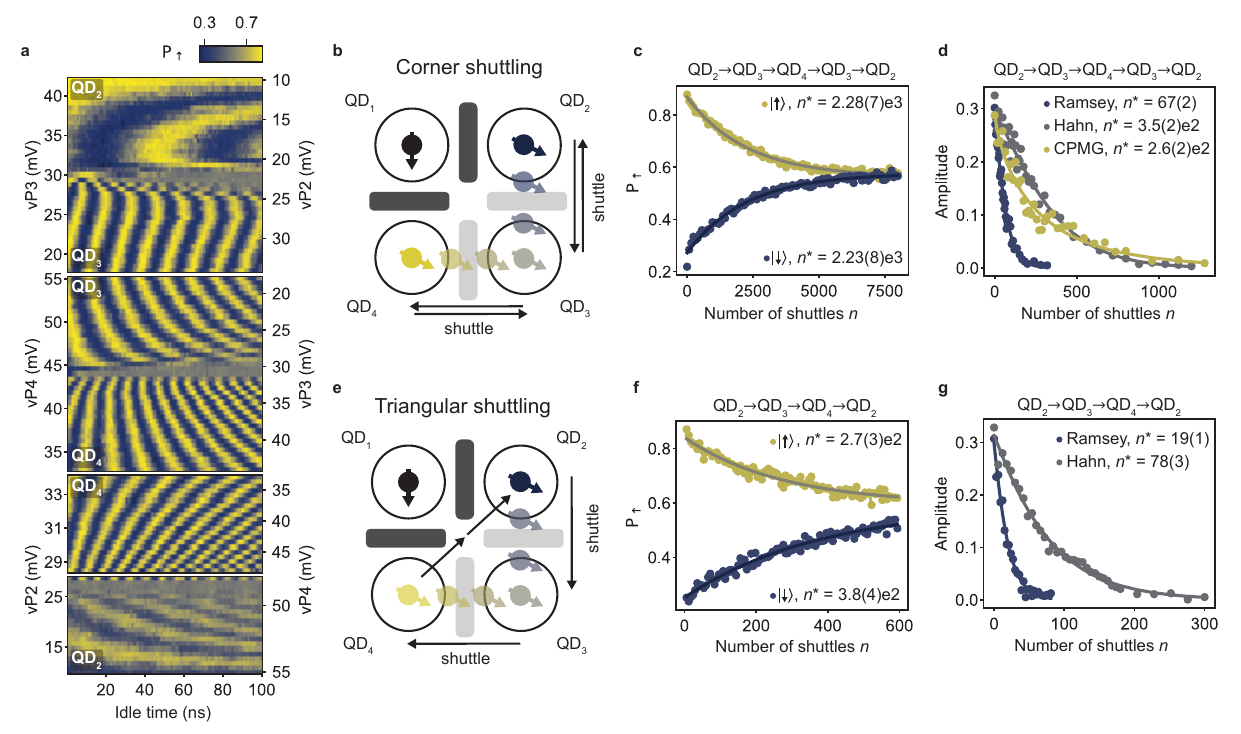}
 \caption{\textbf{Coherent shuttling through quantum dots.} \textbf{a}, Results of free evolution experiments, similar to those displayed in Fig.~\ref{fig:Fig1}.h and l for the corner and triangular shuttling processes. In these experiments, the amplitude of the detuning pulse is increased in steps, in order to shuttle a qubit from QD$_2$ to QD$_3$ and back (top panel), from QD$_2$ to QD$_3$ to QD$_4$ and back (second panel). The measurement in the third panel is identical to the measurement in the second panel, but the final point in the charge stability diagram is stepped towards the charge degeneracy point between QD$_2$ and QD$_4$. In the bottom panel the qubit is shuttled in a triangular fashion: from QD$_2$ to QD$_3$ to QD$_4$ to QD$_2$. The ramp times for this experiment are chosen in such a way that the shuttling is adiabatic with respect to the changes in quantization axis. \textbf{b},  \textbf{e}, Schematic illustrating the shuttling of a spin qubit around the corner: from QD$_2$ to QD$_3$ to QD$_4$ and back via QD$_3$ (b) and in a triangular fashion: from QD$_2$ to QD$_3$ to QD$_4$ and directly back to QD$_2$ (e). The double arrow from QD$_4$ to QD$_2$ indicates that this pulse is made in two steps, in order for the spin to shuttle via the charge degeneracy point of QD$_4$ - QD$_2$ and avoid crossing charge transition lines. \textbf{c}, \textbf{f}, Performance for the corner shuttling (c) and the triangular shuttling (f) of a qubit prepared in the basis states. \textbf{d}, \textbf{g}, Performance for shuttling a qubit prepared in a superposition state for the corner shuttling (d) and the triangular shuttling (g) and for different shuttling sequences. Shuttling performance for different processes are summarized in Supplementary Table 1. }
\label{fig:Fig4}
\end{figure*}

\section*{Shuttling through quantum dots}
 
For distant qubit coupling, it is essential that a qubit can be coherently shuttled through a series of quantum dots. This is more challenging, as it requires control and optimization of a larger amount of parameters. We perform two types of experiments to probe the shuttling through a quantum dot, labelled corner shuttling and triangular shuttling. Fig.~\ref{fig:Fig4}.b shows a schematic of the corner shuttling, which consists of transferring a qubit from QD$_2$ to QD$_3$ to QD$_4$ and back along the same route. The triangular shuttling, depicted in Fig.~\ref{fig:Fig4}.e, consists of shuttling the qubit from QD$_2$ to QD$_3$ to QD$_4$, and then directly back to QD$_2$, without passing through QD$_3$ (for the charge stability diagram QD$_4$-QD$_2$ and a detailed description see Supplementary Note~4).

To probe the feasibility of shuttling through a quantum dot, we measure the free evolution of a coherent state while varying the detuning between the respective quantum dots. The results are shown in Fig~\ref{fig:Fig4}.a. We find a remarkably clear coherent evolution for hole spin transfer from QD$_2$ to QD$_3$ to QD$_4$ and to QD$_2$. We observe one sharp change in the oscillation frequency for each transfer to the next quantum dot. We also note that after completing one round of the triangular shuttling, the phase evolution becomes constant, in agreement with a qubit returning to its original position. We thereby conclude that we can shuttle through quantum dots as desired.

We now focus on quantifying the performance of shuttling through quantum dots by repeated shuttling experiments. To allow comparisons with previous experiments, we define $n$ as the number of shuttling steps between two quantum dots. Meaning that one cycle in the corner shuttling experiments results in $n=4$, while a loop in triangular shuttling takes $n=3$ steps. The results for shuttling basis states are shown in Fig.~\ref{fig:Fig4}.c and ~\ref{fig:Fig4}.f. We note that the spin polarization decays faster compared to the shuttling in double quantum dots, in particular for the triangular shuttling. The corresponding fidelities per shuttling step are $F\simeq99.96~\%$ for the corner shuttling and $F\geq99.63~\%$ for the triangular shuttling.

For the corner shuttling, the faster decay of the basis states suggests a slight increase of the systematic error per shuttling. This may originate from the use of a more elaborated pulse sequence, which makes pulse optimization more challenging. Nonetheless, the characteristic decay constant $n^{\ast}$ remains above 2000 and corresponds to effective distances beyond 300 $\upmu$m (taking a 140~nm quantum dot spacing). The fast decay for the triangular shuttling is likely originating from the diagonal shuttling step. The tunnel coupling between QD$_{2}$ and QD$_{4}$ is low and more challenging to control, due to the absence of a dedicated barrier gate. The low tunnel coupling demands slower ramp times ($t_{\rm ramp}\simeq$ 36~ns) for the hole transfer. This increases the time spent close to the (1,1,0,0)-(1,0,0,1) charge degeneracy point where spin randomization induced by excitations to higher energy states is enhanced~\cite{Krzywda2021}.

Remarkably, we find that the performance achieved for coherent corner shuttling (as shown in Fig.~\ref{fig:Fig4}.d) are comparable to those of coherent shuttling between neighbouring quantum dots. This stems from the performance being limited by dephasing. However, the performance for the CPMG sequence appears inferior when compared to the single echo-pulse sequence. Since the shuttling sequence becomes more complex,  we speculate that it is harder to exactly compensate for the change in quantization axes. Imperfect compensation may introduce transversal noise, which is not fully decoupled using the CPMG sequence. Moreover, close to the anticrossing, the spin is subject to high frequency noise~\cite{Krzywda2021}, whose effect is not corrected and can be enhanced depending on the dynamical decoupling sequence. 

The performance of the coherent triangular shuttling, displayed in Fig.~\ref{fig:Fig4}.g, fall short compared to the corner shuttling. Yet, the number of shuttles reached remains limited by dephasing as shown by the large improvement of $n^{\ast}$ obtained using dynamical decoupling. The weaker performance are thus predominantly a consequence of the use of longer voltage ramps. A larger number of coherent shuttling steps may be achieved by increasing the diagonal  tunnel coupling, which could be obtained by incorporating dedicated barrier gates.

\section*{Conclusion}

We have demonstrated coherent spin qubit shuttling through quantum dots. While holes in germanium provide challenges due to an anisotropic $g$-tensor, we find that spin basis states can be shuttled $n^{\ast}=2230$ times and coherent states up to $n^{\ast}=67$ times and even up to $n^{\ast}=350$ times when using echo pulses. The small effective mass and high uniformity of strained germanium allow for a comparatively large quantum dot spacing of 140~nm. This results in effective length scales for shuttling basis states of $l_{\rm spin}=312~\upmu$m and for coherent shuttling of $l_{\rm coh}=9~\upmu$m. By including echo pulses we can extend the effective length scale to $l_{\rm coh}=49~\upmu$m. These results compare favourably to effective lengths obtained in silicon~\cite{Yoneda2021,Noiri2022,Zwerver2023, Struck2023}. We note that using effective lengths to predict the performance of practical shuttling links requires caution, as the spin dynamics will dependent on the noise of the quantum dot chain. For example, if the noise is local, echo pulses may proof less effective. However, in that case, motional narrowing may facilitate the shuttling~\cite{Huang2013,Flentje2017,Mortemousque2021b,Langrock2022,Struck2023}. Furthermore, operating at even lower magnetic fields and exploiting purified germanium will boost the coherence time and thereby the ability to coherently shuttle. 

%Moreover, we anticipate that the variation in quantization axis between quantum dots and the ability to coherently shuttle between quantum dots will find further applications in qubit definition or characterization.

While we have focused on bucket-brigade-mode shuttling, our results also open the path to conveyor-mode shuttling in germanium, where qubits would be coherently displaced in propagating potential wells using shared gate electrodes. This complementary approach holds promise for making scalable mid-range quantum links and has recently been successfully investigated in silicon~\cite{Struck2023}, though on limited length scales. However, for holes in germanium the small effective mass and absence of valley degeneracy will be beneficial in conveyor-mode shuttling.

Importantly, quantum links based on shuttling and spin qubits are realized using the same manufacturing techniques. Their integration in quantum circuits may provide a path toward networked quantum computing.

%The development of quantum buses based on both shuttling strategies may enable the development of architectures for fault-tolerant quantum computing with spin qubits and on-chip distributed quantum computing.

% Yet, the coherent shuttling still needs to be improved to allow for on-chip electronics~\cite{vandersypen2017,Boter2022}.

% we anticipate that the variation in quantization axis between quantum dots and the ability to coherently shuttle between quantum dots will find further applications in qubit definition, characterization, and development of architectures for fault-tolerant quantum computing. Moreover, the development of a quantum bus based on coherent shuttling, suggests that a monolithic approach could be sufficient to develop networked quantum computing hosting qubits, control and quantum links all on a single semiconductor chip.

\section*{Methods}

\textbf{Materials and device fabrication} 

The device is fabricated on a strained Ge/SiGe heterostructure grown by chemical vapour deposition \cite{Sammak2019,Lodari2019}. From bottom to top the heterostructure is composed of a 1.6~$\upmu$m thick relaxed Ge layer, a 1$~\upmu$m step graded Si$_{1-x}$Ge$_x$ ($x$ going from 1 to 0.8) layer, a 500~nm relaxed Si$_{0.2}$Ge$_{0.8}$ layer, a strained 16~nm Ge quantum well, a 55~nm Si$_{0.2}$Ge$_{0.8}$ spacer layer and a $<1$~nm thick Si cap. 
Contacts to the quantum well are made by depositing 30~nm of aluminium on the heterostructure after etching of the oxidized Si cap. The contacts are isolated from the gate electrodes using a 10~nm aluminium oxide layer deposited by atomic layer deposition. The gates are defined by depositing Ti/Pd bilayers. They are separated  from the each other and from the substrate by 7~nm of aluminium oxide.

\textbf{Experimental procedure} 

To perform the experiments presented, we follow a systematic procedure composed of several steps. We start by preparing the system in a (1,1,1,1) charge state with the hole spins in QD$_1$ and QD$_2$ initialized in a $\ket{\downarrow}$ state, while the other spins are randomly initialized. Subsequently, QD$_3$ and QD$_4$ are depleted to bring the system in a (1,1,0,0) charge configuration. After that, the virtual barrier gate voltage vB$_{12}$ is increased to isolate the ancilla qubit in QD$_1$. The tunnel couplings between QD$_2$ and QD$_3$ and, depending on the experiment, between QD$_3$ and QD$_4$ are then increased by lowering the corresponding barrier gate voltages on vB$_{23}$ and vB$_{34}$. This concludes the system initialization.

Thereafter, the shuttling experiments are performed. Note that to probe the shuttling between QD$_3$ and QD$_4$, the qubit is first transferred adiabatically (with respect to the change in quantization axis) from QD$_2$ to QD$_3$. To determine the final spin state after the shuttlings, the qubit is transferred back adiabatically to QD$_2$. Next, the system is brought back in the (1,1,1,1) charge state, the charge regime in which the readout is optimized. This is done by first increasing vB$_{34}$ and vB$_{34}$, then decreasing vB$_{12}$ and finally reloading one hole in both QD$_3$ and QD$_4$. We finally readout the spin state via latched Pauli spin blockade by transferring the qubit in QD$_1$ to QD$_2$ and integrating the signal from the charge sensor for 7~$\upmu$s. Spin-up probabilities are determined by repeating each experiment a few thousand times (typically 3000). Details about the experimental setup can be found in ref.~\cite{Hendrickx2021}.

\textbf{Achieving sub nanosecond resolution on the voltage pulses} 

The voltage pulses are defined as a sequence of ramps with high precision floating point time stamps and voltages. The desired gate voltage $V(t)$ sequence is generated numerically, sampled at 1 GSa/s (maximum rate achievable with our setup) and then applied on the sample using arbitrary wave form generators (AWGs). To increase the resolution despite the finite sampling rate, we shift the ramps on the desired gate voltage sequence by fractions of nanoseconds. Shifting a ramp by $\tau$ results in a shift of the voltages by $-\tau\frac{\text{d}V(t)}{\text{d}t}$. The AWGs outputting the voltage ramp have a higher order low-pass filter with a cut-off frequency of approximately 400~MHz that smoothens the output signal and effectively removes the effect of the time discretization. The time shift of a pulse is not affected by the filter as the time shift does not change the frequency spectrum of the pulse. Thus the voltage sequence effectively generated on the sample is only delayed by  $\tau$ allowing to achieve a sub nanosecond resolution.

\section*{Acknowledgements}

We thank A. M. J. Zwerver, M. de Smet, L. M. K. Vandersypen, V. V. Dobrovitski and all the members of the Veldhorst group for inspiring discussions. M.V. acknowledges support through two projectruimtes and a Vidi grant, associated with the Netherlands Organization of Scientific Research (NWO), and an ERC Starting Grant. Research was sponsored by the Army Research Office (ARO) and was accomplished under Grant No. W911NF- 17-1-0274. The views and conclusions contained in this document are those of the authors and should not be interpreted as representing the official policies, either expressed or implied, of the Army Research Office (ARO), or the U.S. Government. The U.S. Government is authorized to reproduce and distribute reprints for Government purposes notwithstanding any copyright notation herein. This work is part of the ’Quantum Inspire – the Dutch Quantum Computer in the Cloud’ project (with project number [NWA.1292.19.194]) of the NWA research program ’Research on Routes by Consortia (ORC)’, which is funded by the Netherlands Organization for Scientific Research (NWO).

% We thank O. W. B. Benningshof and R. N. Schouten for technical support.
% \section*{Author contributions}

\section*{Data Availability}
Data supporting this work are available on a Zenodo repository at https://doi.org/10.5281/zenodo.8214452.

\section*{Competing interests}
The authors declare no competing interests. Correspondence should be sent to M. V. (M.Veldhorst@tudelft.nl).

% \section*{Data availability}
% All data underlying this study are available on a online repository at \textbf{\CD{XXX}}.

% \nocite{*}

\bibliography{bib_shuttling}

\clearpage

\end{document}

\title{Supplementary Material: Coherent spin qubit shuttling through germanium quantum dots}

\author{Floor van Riggelen-Doelman}
\affiliation{QuTech and Kavli Institute of Nanoscience, Delft University of Technology, PO Box 5046, 2600 GA Delft, The Netherlands}
\author{Chien-An Wang}
\affiliation{QuTech and Kavli Institute of Nanoscience, Delft University of Technology, PO Box 5046, 2600 GA Delft, The Netherlands}
\author{Sander L. de Snoo}
\affiliation{QuTech and Kavli Institute of Nanoscience, Delft University of Technology, PO Box 5046, 2600 GA Delft, The Netherlands}
\author{William I. L. Lawrie}
\affiliation{QuTech and Kavli Institute of Nanoscience, Delft University of Technology, PO Box 5046, 2600 GA Delft, The Netherlands}
\author{Nico W. Hendrickx}
\affiliation{QuTech and Kavli Institute of Nanoscience, Delft University of Technology, PO Box 5046, 2600 GA Delft, The Netherlands}
\author{Maximilian Rimbach-Russ}
\affiliation{QuTech and Kavli Institute of Nanoscience, Delft University of Technology, PO Box 5046, 2600 GA Delft, The Netherlands}
\author{Amir Sammak}
\affiliation{QuTech and Netherlands Organisation for Applied Scientific Research (TNO), Delft, The Netherlands}
\author{Giordano Scappucci}
\affiliation{QuTech and Kavli Institute of Nanoscience, Delft University of Technology, PO Box 5046, 2600 GA Delft, The Netherlands}
\author{Corentin Déprez}
\affiliation{QuTech and Kavli Institute of Nanoscience, Delft University of Technology, PO Box 5046, 2600 GA Delft, The Netherlands}
\author{Menno Veldhorst}
\affiliation{QuTech and Kavli Institute of Nanoscience, Delft University of Technology, PO Box 5046, 2600 GA Delft, The Netherlands}
\date{\today}
\maketitle

\setcounter{table}{0}
\setcounter{figure}{0}
\setcounter{section}{0}

\makeatletter 
\makeatother
\onecolumngrid
\begin{center}
\vspace{-1.5cm}

% \Letter{~:~\href{mailto:C.C.Deprez@tudelft.nl}{C.C.Deprez@tudelft.nl},~\href{mailto:M.Veldhorst@tudelft.nl}{M.Veldhorst@tudelft.nl}}
\end{center}
\vspace{0.5 cm}
This Supplementary Material includes :

\begin{itemize}
    \item Supplementary Notes 1-6
    \item Supplementary Figures 1-8
    \item Supplementary Table 1
    \item Supplementary References 1-8
\end{itemize}
\clearpage
\vspace{3cm}
\begin{figure}[h!]
\centering
\includegraphics[width=1\textwidth]{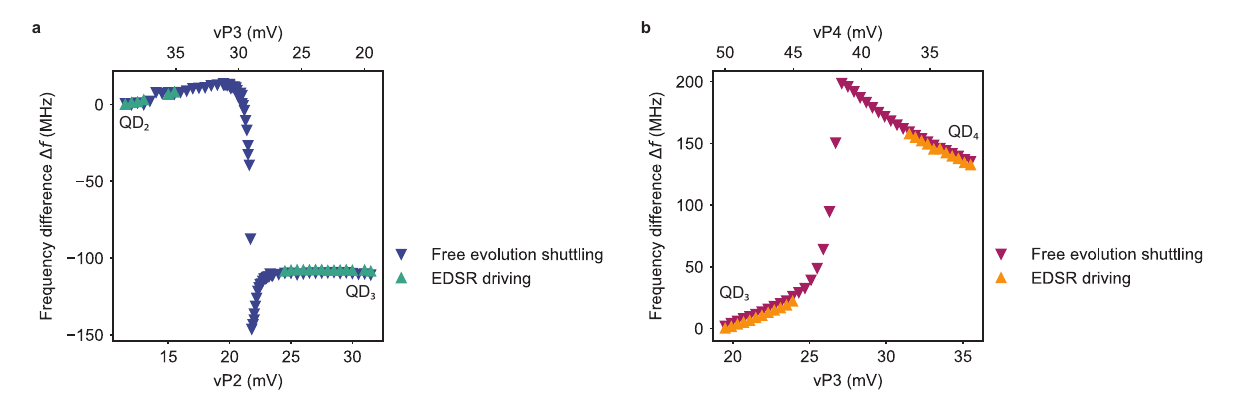}
\caption{\textbf{Evolution of the Larmor frequency for shuttling in double quantum dots. a}, \textbf{b}, Larmor frequency differences $\Delta f=f_{\rm L}(\text{vP}3)-f^{\rm QD2}_{\rm L}$ (a)  and $\Delta f=f_{\rm L}(\text{vP}4)-f^{\rm QD3}_{\rm L}$ (b) measured along the detuning axis of QD$_{2}$-QD$_{3}$ (a) and QD$_{3}$-QD$_{4}$ (b). The quantum dot where the shuttling experiment starts is taken as the reference point for the frequency. $\Delta f$ is independently evaluated from measurements of the resonance frequency using an EDSR pulse (data displayed in Fig.~1.g and k) and from the frequency of the coherent oscillations that appear when a qubit is shuttled in a superposition state (data displayed in Fig.~1.h and l). Both sets of data points overlap in (a) and (b), confirming that coherent oscillations arise due to a change in Larmor frequency along the detuning axis. For the free evolution experiments, the shuttling between QD$_{2}$ and QD$_{3}$ (shown in (a)) is completely adiabatic (ramp times of 40~ns) while the shuttling between QD$_{3}$ and QD$_{4}$ (shown in (b)) is only partially adiabatic (ramp times of 4~ns). In the latter case, the frequency difference measured is barely affected by the limited adiabaticity as the visibility $M$ of the oscillations induced by the change in quantization axis ($M<0.1$ from \ref{figSup:Tilt_angle_from_amplitude}) is sufficiently small compared to that of the oscillations arising from the phase evolution of the superposition state ($V\approx0.5$ when the hole is in QD$_4$). Moreover, the Larmor frequency of both a spin in QD$_3$ and in QD$_4$ is very close to 1~GHz. The free evolution experiments were performed with 1~ns time precision, meaning that the oscillations due to the diabaticity of the shuttling only show up as an aliasing pattern and do not disturb the oscillations due to free evolution. }
\label{figSup:Res_Freq_compare_to_free_evolution}
\end{figure}

\vspace{3cm}
\begin{center}   

\begin{table*}[h!]
\begin{center}
\begin{tabular}{|c|c|c|c|c|} 
 \hline
 Shuttling process & $n^{\ast}$ for $\ket{\downarrow}$ transfer & $n^{\ast}$ for $\ket{\uparrow}$ transfer& $n^{\ast}$ for $\frac{\ket{\downarrow}+i\ket{\uparrow}}{\sqrt{2}}$ transfer & $\alpha$ for $\frac{\ket{\downarrow}+i\ket{\uparrow}}{\sqrt{2}}$ transfer    \\ [0.5ex] 
 \hline\hline
&  &    & Ramsey: $64\pm1$ &  Ramsey: 1.36$\pm0.05$\\
QD$_{2} \rightleftarrows$ QD$_3$ & 3.36$\times10^{3}\pm90$  &  3.2$\times10^{3}\pm100$  & Hahn: $376\pm5$ & Hahn: 1.44$\pm0.04$\\
& & & CPMG: $450\pm20$ & CPMG: 1.14$\pm0.06$\\
\hline
&  &    & Ramsey: $77\pm2$ &  Ramsey: 1.28$\pm0.06$\\
QD$_{3} \rightleftarrows$ QD$_4$ & 2.9$\times10^{3}\pm100$  &  3.1$\times10^{3}\pm100$  & Hahn: $332\pm6$ & Hahn: 1.17$\pm0.04$\\
& & & CPMG: $500\pm10$ & CPMG: 1.3$\pm0.07$\\
\hline
Corner &  &    & Ramsey: $67\pm2$ &  Ramsey: 1.11$\pm0.06$\\
QD$_{2} \rightarrow$ QD$_3\rightarrow$ QD$_4$ & 2.23$\times10^{3}\pm80$  &  2.28$\times10^{3}\pm70$  & Hahn: $350\pm20$ & Hahn: 1.2$\pm0.1$ \\
$\rightarrow$ QD$_{3} \rightarrow$ QD$_2$& & & CPMG: $260\pm20$ & CPMG: 0.76$\pm0.07$ \\
\hline
Triangular &  &    & Ramsey: $19\pm1$ &  Ramsey: 1.08$\pm0.07$\\
QD$_{2} \rightarrow$ QD$_3 \rightarrow$ QD$_4 \rightarrow $QD$_{2}$ & 380$\pm40$  &  270$\pm30$  & Hahn: $78\pm3$ & Hahn: 1.07$\pm0.05$ \\
\hline
 \end{tabular}
\end{center}
\caption{\textbf{Summary of shuttling performance.} For the spin basis state shuttling experiments, the spin polarization decays with the number of shuttles $n$ are fitted by $\text{P}_0\exp(-(n/n^{\ast}))+\text{P}_{\rm sat}$.  For the coherent shuttling experiments, the coherence decays are fitted by $A_0\exp{(-({n/n^{\ast}})^\alpha)}$. $n^{\ast}$ represents the number of shuttles that can be achieved before the polarization the coherence or drops by 1/e.}
\label{tableSup:Table1}
\end{table*}
\end{center}  

\clearpage

\section{Q\MakeLowercase{ubit resonance frequency nearby the interdot charge transition}}

In Fig.~1.g and k, we show the evolution of the qubit resonance frequency $f_{\rm L}$ along the detuning axis of the QD$_{2}$-QD$_{3}$ quantum dot pair and of the QD$_{3}$-QD$_{4}$ quantum dot pair. $f_{\rm L}$ is measured by shuttling the spin and applying a 4~$\upmu$s long EDSR pulse on one plunger gate. While  $f_{\rm L}$ can be clearly determined when the hole is well-localized in one quantum dot, it cannot be measured nearby the charge transition as the spin-up probability has a high value over the whole range of frequency spanned. We think that this is the result of a combination of different effects.

Since the two quantum dots have different quantization axes, the system effectively behaves as a flopping-mode qubit nearby the charge transition~\cite{Benito2019,Croot2020,Mutter2021b,Hu2023} and the EDSR driving is thus expected to be more efficient. This appears, in Fig.~1.g, when the qubit is in QD$_2$: along the resonance line, we observe an alternation of high and low spin-up probabilities that witness rapid variations of the Rabi frequency. As a consequence, the power broadening increases significantly in the vicinity of the charge transition which prevents us from resolving the qubit resonance frequency. Likewise, the gradient of shear strains induced by the thermal contraction of the gate electrodes can lead to large increases of the Rabi frequency~\cite{Abadillouriel2022}. It is likely that this effect is enhanced in the vicinity of the charge transition, as the hole is delocalized between the two quantum dots and its wavefunction extends below the edges of several gates. Finally, nearby the charge transition, excitations to higher energy states induced by charge noise are more likely to occur~\cite{Krzywda2021}, especially on the relatively long timescale of 4~$\upmu$s. These transitions to higher energy states lead to a randomization of the spin states, which explain the large spin-up probabilities observed over the full frequency range. 

\section{Q\MakeLowercase{uantifying the quantization axis tilt angle}}

\subsection{E\MakeLowercase{stimation based on the visibility of the oscillations induced by the change in quantization axis}}

The tilt angle $\theta$ between the quantization axis of two different quantum dots can be estimated based on the amplitude of the oscillations induced by diabatically shuttling a qubit in the $\ket{\downarrow}$ state. This approximation relies on a simple geometric construction in the Bloch sphere.

\begin{figure}[h!]
\centering
\includegraphics[width=1\textwidth]{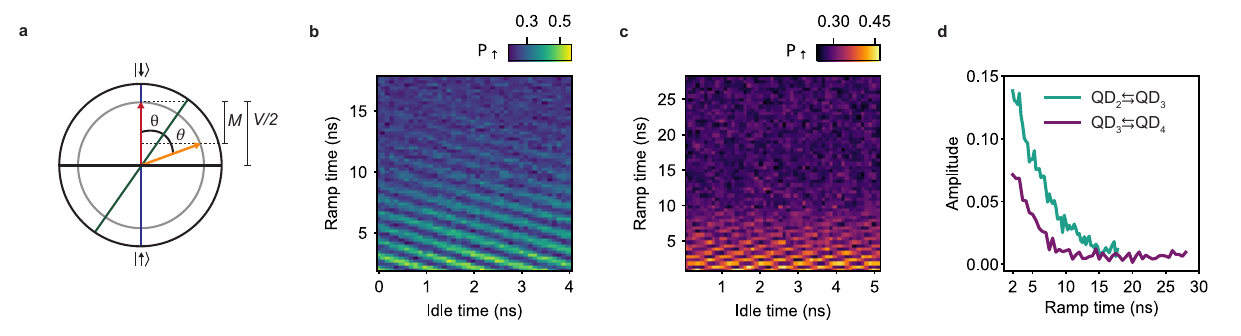}
\caption{\textbf{Estimation of the tilt angle based on the amplitude of the oscillations induced by the difference in quantization axis. a,} Geometric construction in the Bloch sphere allowing to determine the tilt angle $\theta$ between the quantization axes of adjacent quantum dots (blue and dark green). $\theta$ is determined from the visibility $M$ of the oscillations induced by the change in quantization axes and the visibility of the Rabi oscillations $V$. \textbf{b}, \textbf{c}, Oscillations induced while shuttling a qubit in a $\ket{\downarrow}$ state between QD$_{2}$ and QD$_{3}$ (b) and between QD$_{3}$ and QD$_{4}$ (c) for increasing ramp times. \textbf{d}, Amplitude of the oscillations as function of the ramp times.}
\label{figSup:Tilt_angle_from_amplitude}
\end{figure}

\ref{figSup:Tilt_angle_from_amplitude}.a shows the Bloch sphere projected on the plane defined by the quantization axes of the two quantum dots (dark blue and dark green). At the beginning of the experiment, the qubit is initialized in the $\ket{\downarrow}$ state (red arrow). After shuttling to the neighboring quantum dot, the qubit state changes due to the difference between the quantization axes. In the Bloch sphere, it can be represented by rotations of the state vector around the second quantization axis. After half a period (orange arrow), the state projection on the quantization axis of the quantum dot where the experiment started differs maximally from that of the initial state. This sets the visibility $M$ of the oscillations induced by the change of quantization axis. 

In practise, this visibility is reduced due to imperfect initialization and readout. This can be taken into account by assuming that the state vectors have a norm $V/2<0.5$ with $V$ being the visibility of Rabi oscillations measured in the quantum dot where the shuttling experiment starts. We neglect relaxation which is irrelevant at the time scale of few nanoseconds~\cite{Lawrie2020} and thus assume that the norm of the vector state stays constant during the rotations. We find that: 
\begin{equation}
    \theta=\frac{1}{2}\arccos(1-2M/V) \text{ with } 0\leq\theta\leq\pi.
    \label{eq:angle_amplitude}
\end{equation}

We use this expression to evaluate $\theta_{23}$ ($\theta_{34}$), the tilt angle between the quantization axes of QD$_{2}$ and QD$_{3}$ (QD$_{3}$ and QD$_{4}$). \ref{figSup:Tilt_angle_from_amplitude}.b and c show the amplitudes $M/2$ of the oscillations induced by the change in quantization axis as function of the pulse ramp time $t_{\rm ramp}$. As discussed in the main text, the amplitudes drop rapidly to zero as $t_{\rm ramp}$ increases, because the shuttling becomes more adiabatic with respect to the difference in quantization axis. For the evaluation of $\theta$ we use the amplitude $M/2=0.14$ (0.07) of the oscillations at the shortest $t_{\rm ramp}=2$~ns. We remark that there is no clear saturation of $M$ at the smallest ramp times, which suggests that the shuttling process is still not fully diabatic and that higher visibilities could be achieved by shuttling faster. Rabi oscillations for the driving of the qubit in QD$_2$ (QD$_3$) have a visibility of $V=0.61$ (0.48) giving us $\theta_{23}\geq42\degree$ ($\theta_{34}\geq33\degree$). These large values for $\theta$ illustrate the strong influence of the local electric field on the direction of the quantization in germanium hole spin qubits operated with an in-plane external magnetic field.

%and for different voltages applied on the barrier gate vB$_{23}$ (vB$_{34}$)

% Function used to estimate the difference in quantization axis between two quantum dots: $\theta = 0.5 * arccos (2(0.5V - M)/V )$. 
% Figure a: shuttling from QD$_{2}$ to QD$_{3}$, maximum amplitude in oscillations (2-2.5 ns): 0.13, amplitude of Rabi oscillation of spin qubit in QD$_{2}$: 0.31, estimated lower limit of the difference in quantization between QD$_{2}$ and QD$_{3}$: 42 degrees.
% Figure b:shuttling from QD$_{3}$ to QD$_{4}$, maximum amplitude in oscillations (2-2.6 ns): 0.07, amplitude of Rabi oscillation of spin qubit in QD$_{3}$: 0.24, estimated lower limit of the difference in quantization between QD$_{3}$ and QD$_{4}$: 33 degrees.

\subsection{Estimations based on fits with a four-level model}

% The tilt angle can also be evaluated from the evolution of the qubit resonance along the detuning axis.

To get additional independent evaluation of the tilt angles, we can fit the evolution of the qubit resonance with a four-level model. To derive such a model, we consider a single hole in a germanium double quantum dot placed in an external magnetic field $B$. We assume that there is a finite tunnel coupling $t_{\rm c}$ between the two quantum dots QD$_{\rm A}$ and QD$_{\rm B}$ and their quantization axis are tilted with respect to each other by an angle $\theta$. This last assumption is sufficient to take into account all effects of the spin-orbit interaction, providing a suitable basis transformation and a renormalization of the tunneling terms.

The system can then be described in the basis $\{\ket{\rm A,\uparrow_{\rm A}},\ket{\rm A,\downarrow_{\rm A}},\ket{\rm B,\uparrow_{\rm A}},\ket{\rm B,\downarrow_{\rm A}}\}$, where `A' or `B' indicates the position of the hole in quantum dot QD$_{\rm A}$ or QD$_{\rm B}$ and $\uparrow_{\rm A}$ or $\downarrow_{\rm A}$ specifies its spin states in the frame of quantum dot A. Its Hamiltonian is then given by:

\begin{equation}
H_{\rm model} = H_{\rm charge} +H_{\rm Zeeman}=
  \left( {\begin{array}{cccc}
     \epsilon & 0 & t_{\rm c} & 0 \\
   0 & \epsilon & 0 & t_{\rm c}\\
   t_{\rm c} &  0 & -\epsilon & 0 \\
   0 &  t_{\rm c} & 0 & -\epsilon \\
  \end{array} } \right) +  \frac{1}{2}B\mu_{\rm B}\left( {\begin{array}{cccc}
    g_{\rm A}(\epsilon) & 0 & 0 & 0\\
   0 & -g_{\rm A}(\epsilon) & 0 & 0\\
   0& 0 & g_{\rm B}(\epsilon)\cos(\theta)& g_{\rm B}(\epsilon)\sin(\theta)\text{e}^{\text{i}\varphi}\\
  0 & 0 & g_{\rm B}(\epsilon)\sin(\theta)\text{e}^{-\text{i}\varphi}& -g_{\rm B}(\epsilon)\cos(\theta)\\
  \end{array} } \right),
\label{Hamiltonian_tilt_angle}
\end{equation}
where $\epsilon$ is the detuning energy of the double quantum dot system (taken as zero at the charge transition), $\mu_{\rm B}$ is the Bohr magneton and the $g_{i}$ are the $g$-factors in the different quantum dots, $\varphi$ is the azimuthal angle between the two quantization axes. We note that this model is similar to that of a flopping-mode qubit~\cite{Benito2019}. Diagonalizing the Hamiltonian, we obtain the qubit resonance frequency $f_{\rm L}$ given by:

\begin{equation}
f_{\rm L} =   \frac{\mu_{\rm B}B}{h}\frac{\sqrt{(2\epsilon^2+t_{\rm c}^2)(g_{\rm A}(\epsilon)^2+g_{\rm B}(\epsilon)^2)+2\epsilon(g_{\rm B}(\epsilon)^2-g_{\rm A}(\epsilon)^2)\sqrt{\epsilon^2+t_{\rm c}^2}+2g_{\rm A}(\epsilon)g_{\rm B}(\epsilon)t_{\rm c}^2\cos(\theta)}}{2\sqrt{\epsilon^2+t_{\rm c}^2}},
\label{Res_freq_AC}
\end{equation}

% The evolution of $f_{\rm L}$ along the detuning axes can then be fitted to extract the tilt angles and the tunnel couplings between neighbouring quantum dots. For this purpose, we first express the detuning energies in terms of gate voltages as $\epsilon_{23}=\eta_{23}(\rm vP_{3}-vP_{3}^0)$ and $\epsilon_{34}=\eta_{34}(\rm vP_{4}-vP_{4}^0)$ where $\eta_{23}=0.166$ and $\eta_{34}=0.150$ are the effective detuning lever arm which determined via photon-assisted tunnelling experiments~\cite{Oosterkamp1998}. We then extract the evolution of $f_{\rm L}$ as function of vP$_{3}$ (vP$_{4}$) from the data displayed in~\ref{figSup:Tilt_fit}.a-b (\ref{figSup:Tilt_fit_Q3Q4}.a-c) and fit it with eq.(\ref{Res_freq_AC}).

\begin{figure}[h!]
\centering
\includegraphics[width=1\textwidth]{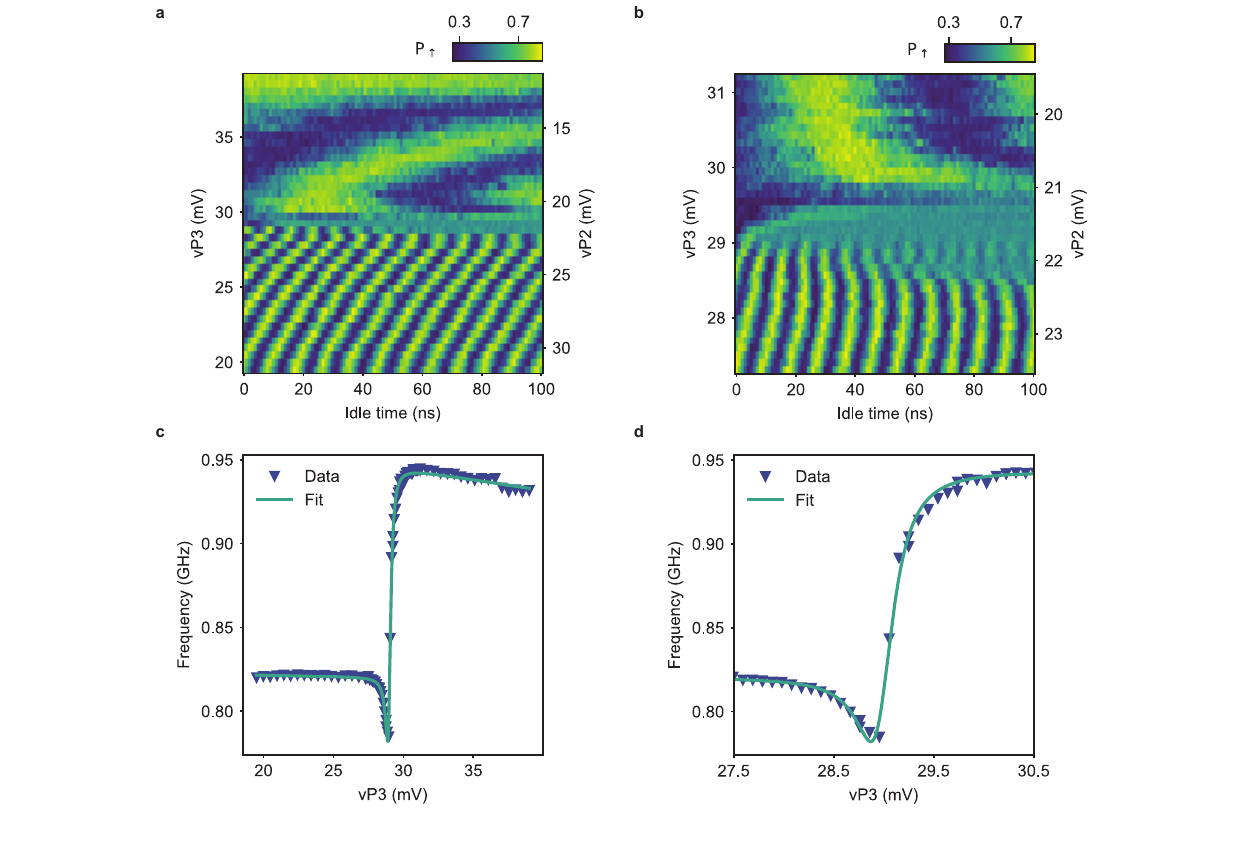}
\caption{\textbf{Evaluation of the tilt angle between QD$\mathbf{_2}$ and QD$\mathbf{_3}$ quantization axes using a four-level model. a}, Free evolution experiments for shuttling a qubit in superposition state between QD$_2$ and QD$_3$ back-and-forth. The superposition state is prepared in  QD$_2$. \textbf{b}, Zoom-in on the vicinity of the charge transition. The two data sets are identical to those displayed in Fig.~1.h. \textbf{c}, \textbf{d}, Resonance frequency extracted from the oscillations along the detuning axis in (a) and (b) and fit with the model of eq.~(\ref{Res_freq_AC}).}
\label{figSup:Tilt_fit}
\end{figure}

\begin{figure}[h!]
\centering
\includegraphics[width=1\textwidth]{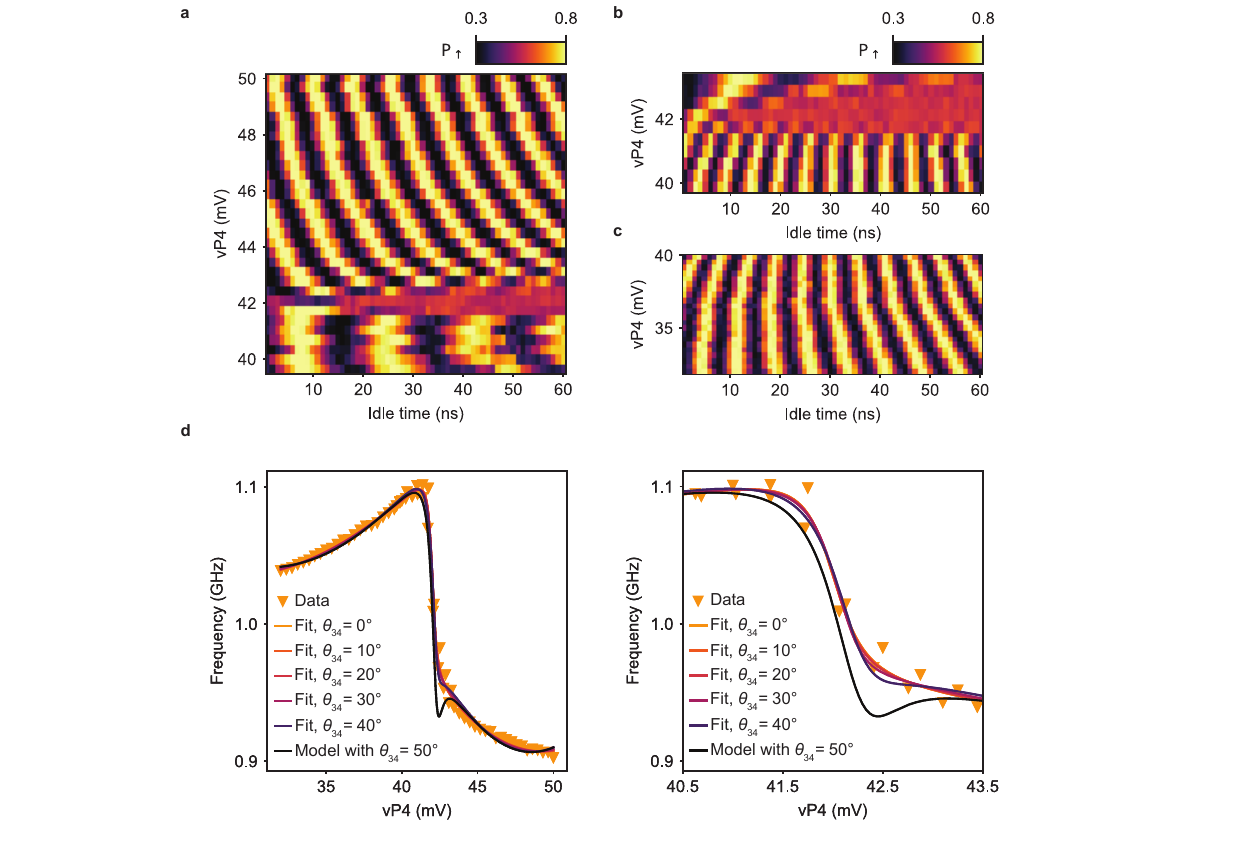}
\caption{\textbf{Evaluation of the tilt angle between QD$\mathbf{_3}$ and QD$\mathbf{_4}$ quantization axes using a four-level model. a}, \textbf{b}, \textbf{c}, Free evolution experiments for the adiabatic shuttling of a qubit in superposition state between QD$_3$ and QD$_4$ back-and-forth. In (a) the qubit is prepared in superposition in QD$_4$, while in (b) and (c) the superposition state is prepared in QD$_3$. \textbf{c},  Evolution of the resonance frequency along the detuning axis, extracted from the oscillations in (a), (b) and (c), and fits with models of eq.~(\ref{Res_freq_AC}) where the tilt angle is fixed. The expected evolution for $\theta_{34}=50\degree$ is computed using the parameters extracted from the fit with $\theta_{34}=40\degree$ }
\label{figSup:Tilt_fit_Q3Q4}
\end{figure}

The evolution of $f_{\rm L}$ along the detuning axes can then be fitted to extract the tilt angles and the tunnel couplings between neighbouring quantum dots. For this purpose, we first express the detuning energies in terms of gate voltages as $\epsilon_{23}=\eta_{23}(\rm vP_{3}-vP_{3}^0)$ and $\epsilon_{34}=\eta_{34}(\rm vP_{4}-vP_{4}^0)$ where $\eta_{23}=0.166$ and $\eta_{34}=0.150$ are the effective lever arms along the detuning axis. They are defined as $\eta_{23}=\beta_{3}+\beta_{2}\gamma_{23}$ and $\eta_{34}=\beta^{\ast}_{4}+\beta^{\ast}_{3}\gamma_{34}$ where $\beta^{(\ast)}_{i}$ are the virtual gate lever arms measured nearby the QD$_2$-QD$_3$ (QD$_3$-QD$_4$) charge transition via photon-assisted tunnelling experiments~\cite{Oosterkamp1998} and where the $\gamma_{ij}=\vert\Delta \text{vP}_i/\Delta \text{vP}_j \vert$ are the slopes of the detuning axis. We then extract the evolution of $f_{\rm L}$ as function of vP$_{3}$ (vP$_{4}$) from the data displayed in~\ref{figSup:Tilt_fit}.a-b (\ref{figSup:Tilt_fit_Q3Q4}.a-c) and fit it with eq.(\ref{Res_freq_AC}).
% assuming a linear (quadratic) dependence of the $g$-factor with the gate voltage

\ref{figSup:Tilt_fit}.c-d display the evolution of $f_{\rm L}$ along the $\epsilon_{23}$ detuning axis which is fitted to the above model assuming a linear dependence of $g$ with vP$_{3}$. We observe that the model reproduces well the measured evolution. This allows to estimate an interdot tunnel coupling $t_{\rm c}$ of $8.7\pm0.3$~GHz and a tilt angle $\theta_{23}$ of $=51.8\pm0.7\degree$. The latter is consistent with the lower bound found using the previous method. 

\ref{figSup:Tilt_fit_Q3Q4}.d-e display the evolution of $f_{\rm L}$ along the $\epsilon_{34}$ detuning axis. In this case, fitting the data does not allow to extract the tilt angle, even if we assume a quadratic dependence of the $g$-factor with the gate voltage. Indeed, for $0\degree\leq\theta\lesssim40\degree$, the shape of $f_{\rm L}$ curve is nearly solely determined by the tunnel coupling and the variation of the $g$-factor with vP$_{4}$. Consequently, the data can be equally well fitted by models where $\theta_{34}$ is fixed 0\degree, 10\degree, 20\degree, 30\degree or 40\degree. This leads to large uncertainty on the value of $\theta_{34}$ that prevents us to extract it. Nevertheless, the tunnel coupling between QD$_3$ and QD$_4$ can still be estimated from these fits and, for $\theta_{34}$ fixed to 40\degree (30\degree), we find $t_{\rm c}=15\pm2$ ($t_{\rm c}=12\pm$2) GHz.

What does become clear, however, is that we cannot obtain proper fits of the data with model where $\theta_{34}$ is fixed to values larger than 40$\degree$. The underlying reason appears when plotting the expected evolution of $f_{\rm L}$ in such model: for $\theta_{34}\gtrsim50$\degree,  $f_{\rm L}$  should display a minimum that we do not observe experimentally. This suggests that $\theta_{34}$ is lower than $50$\degree.

\clearpage

\vspace{-0.5cm}
\section{A\MakeLowercase{diabatic shuttling}}

\begin{figure}[h!]
\centering
\includegraphics[width=1\textwidth]{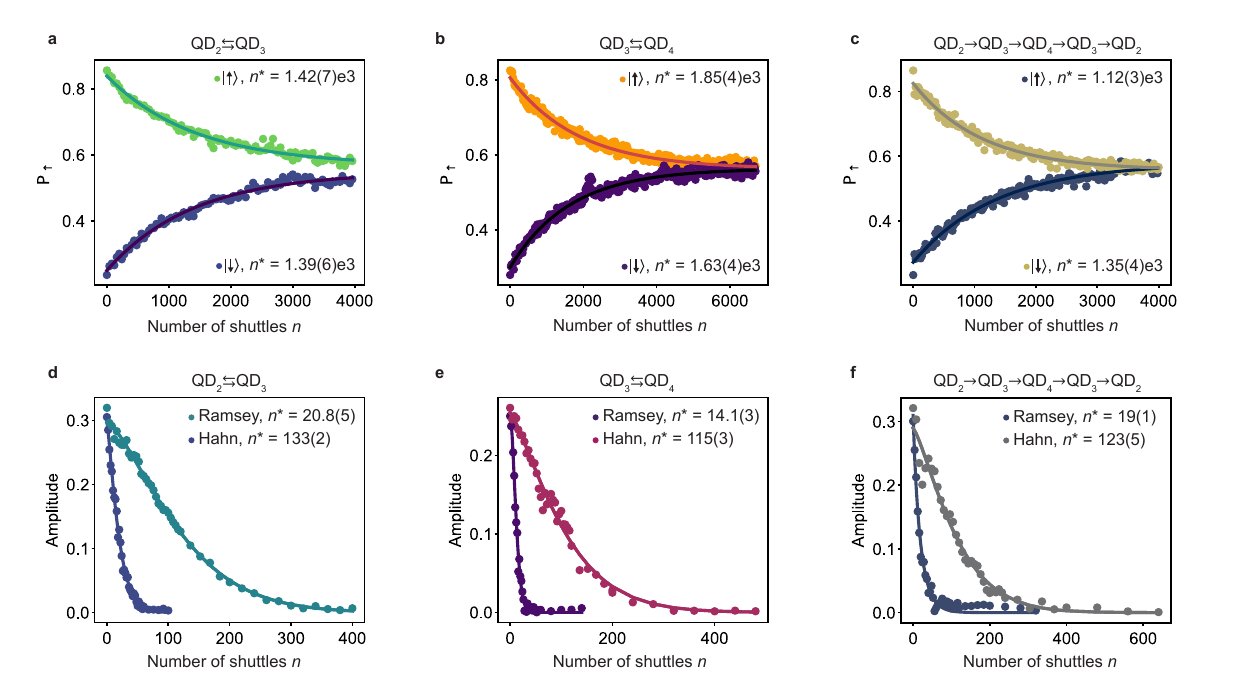}
\caption{\textbf{Performance of adiabatic shuttling. a}, \textbf{b}, \textbf{c}, Spin polarization as a function of the number of shuttling steps $n$ for a qubit initialized in the basis states. \textbf{d}, \textbf{e}, \textbf{f}, Amplitude as a function of the number of shuttling steps $n$ for qubits initialized in a superposition state, without (Ramsey) and with echo pulse (Hahn).}
\label{figSup:Adiabatic_shuttling}
\end{figure}

For completeness, we also investigate the performance of the shuttling processes when the shuttling pulses are adiabatic, i.e. when there is no rotation induced by the difference between the quantization axes of the quantum dots. \ref{figSup:Adiabatic_shuttling} shows the results of such investigations for the shuttling of basis states and for the shuttling of superposition states.  In both cases, we obtain significantly lower performance compared to those achieved with diabatic pulses (see Figure 3 in the main text). According to our findings, dephasing can largely explain this difference in performance for the coherent shuttling experiments. As the time required for each shuttling event is increased in the adiabatic experiments, the qubit experiences more dephasing during each shuttling step and the phase coherence is lost after a smaller number of shuttling steps $n$. The use of echoing pulses allows us to get an improvement of the coherent shuttling performance by a factor 6 to 8, larger than those obtained for diabatic shuttling.

For shuttling of basis states, the lower performance suggests that the probability of having a spin-flip during a shuttling increases if the latter is performed adiabatically. This could originate from the longer time spent in the vicinity of the charge transition, where spin randomization induced by charge noise is enhanced~\cite{Krzywda2021}. Overall, the data in \ref{figSup:Adiabatic_shuttling} clearly show that an approach based on diabatic spin shuttling is preferable for hole spin qubits in germanium.

\section{C\MakeLowercase{harge stability diagram of pair }\MakeUppercase{QD$\mathbf{_2}$-QD$\mathbf{_4}$} \MakeLowercase{and  triangular shuttling}}

\begin{figure}[h!]
\centering
\includegraphics[width=1\textwidth]{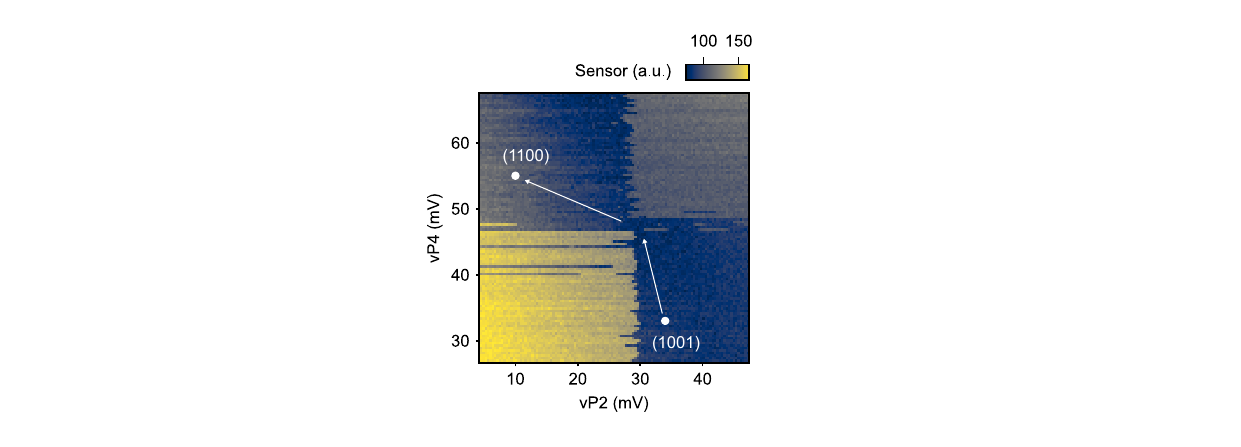}
\caption{\textbf{Charge stability diagram of quantum dot pair QD$\mathbf{_2}$-QD$\mathbf{_4}$.} No clear interdot transition can be distinguished. The shuttling of a spin qubit from QD$_2$ to QD$_4$ is performed using two voltages pulses (white arrows).}
\label{figSup:CSD_triangle}
\end{figure}

The charge stability diagram of the quantum dot pair QD$_2$-QD$_4$, measured in a configuration identical to that of the triangular shuttling, is displayed in \ref{figSup:CSD_triangle}. No clear interdot charge anticrossing is visible, which suggests that the tunnel coupling between the two quantum dots is very low. This is expected, considering the device geometry, and it forces us to split the final pulse for the triangular shuttling in two parts. As depicted in \ref{figSup:CSD_triangle}, the voltages are first changed to bring the system close to the (1,1,0,0)-(1,0,0,1) degeneracy point before applying a second pulse that brings the system in the (1,1,0,0) charge state. This reduces the probability that we excite the (1,1,0,1) charge state, while transferring the qubit.

\section{O\MakeLowercase{ptimization of the shuttling pulses to mitigate the effects of spin-orbit interaction}}

\begin{figure}[h!]
\centering
\includegraphics[width=1\textwidth]{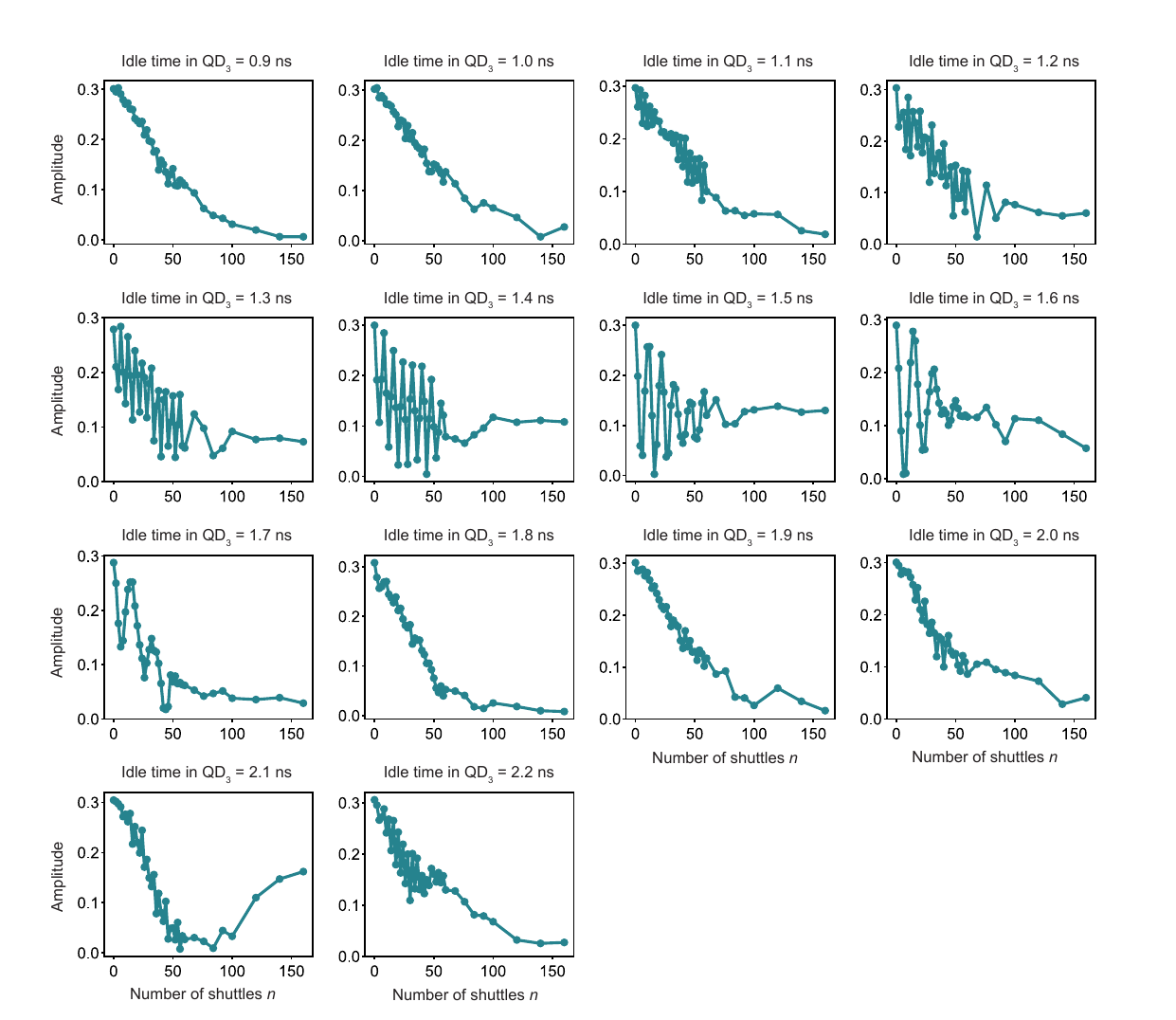}
\caption{\textbf{Signatures of non-optimized idle times in Ramsey shuttling experiments.} Results of coherent shuttling experiments between QD$_2$ and QD$_3$ obtained using Ramsey sequences. The idle time spent in QD$_3$ is different for the results shown in the different subplots, as indicated by the titles. For non-optimized idle times, oscillations of the amplitude as function of the number of shuttles $n$ appear and the amplitude can saturate to a non-zero value at large $n$.}
\label{figSup:vary_waitQ3}
\end{figure}

In this section, we illustrate and discuss the importance of careful pulse optimization. \ref{figSup:vary_waitQ3} shows the results of experiments where we probe the performance of the coherent shuttling between QD$_2$ and QD$_3$ using the Ramsey sequence depicted in Fig.~3.a. The detuning pulses used for all these experiments are identical, except for the idle time $t_{\rm idle}$ in QD$_3$ (idle time 2 in Fig.~3.b). This idle time in QD$_3$ was optimized to 0.95~ns for the experiments displayed in the main text.

We observe that the evolution of amplitudes extracted at the end of the shuttling sequence is strongly dependent on the idle time in QD$_3$. For $t_{\rm idle}=0.9$ and $t_{\rm idle}=1$~ns, which are close to the optimum, the amplitude shows a smooth and progressive decay. When $t_{\rm idle}$ is increased, oscillations of the amplitude as function of the number of shuttling steps $n$ appear and their periodicity varies with $t_{\rm idle}$. These oscillations witness the rotations induced by the change of quantization axes, which are imperfectly compensated for $t_{\rm idle}\geq1.1~$ns. They lead to coherent errors after each shuttling event, which add up, and significantly modify the state of the qubit. For example, for $t_{\rm idle}=1.6~$ns, the superposition state is virtually transformed to a spin basis state after a few shuttling rounds. This emphasizes the necessity of optimizing the voltages pulses to compensate for the effect of rotations induced by the spin-orbit interaction. 

%observed while varying the phase of the second $\pi$/2 pulse

The optimized idle times for the each shuttling processes can be found by performing measurements similar to those displayed in \ref{figSup:vary_waitQ3}, and by looking for regular decay of the amplitude as function of $n$. This optimization can also be done similarly studying the decay of the spin-up probabilities in spin basis state shuttling experiments.

\section{Q\MakeLowercase{ubit dynamics during coherent shuttling experiments for non-optimized idle times}}

\begin{figure}[h!]
\centering
\includegraphics[width=1\textwidth]{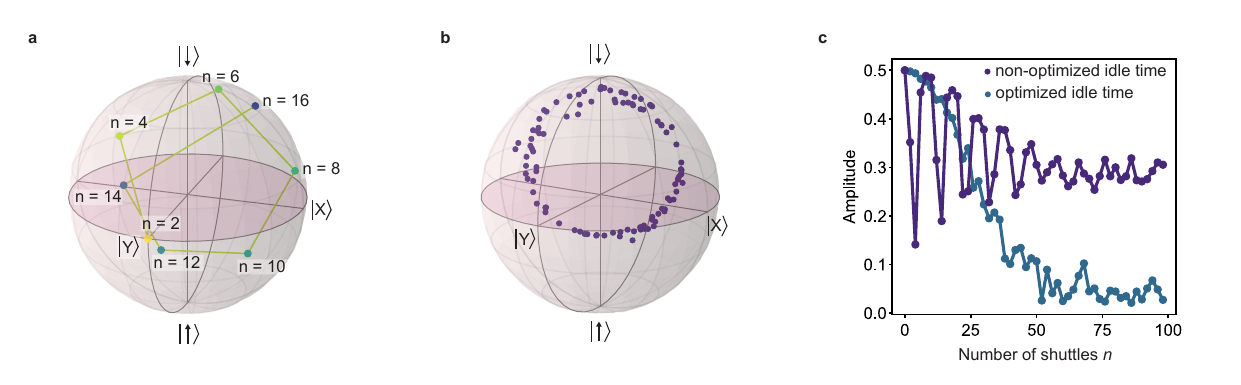}
\caption{\textbf{Simulation of the effect of non-optimized idle times.} \textbf{a}, Distribution of the qubit states after an even number of shuttles, for an non-optimized idle time. \textbf{b}, Spread of the qubit state after a large number of shuttles, when the qubit is dephased. \textbf{c}, Simulated measurement results, i.e. amplitude of the oscillations appearing while varying the phase of the second $\pi/2$ pulse, as a function of $n$, for a non-optimized idle time and an optimized idle time. }
\label{figSup:effect_wrong_waittime}
\end{figure}

In \ref{figSup:vary_waitQ3}, we see that for non-optimized idle times, like $t_{\rm idle}=1.5$~ns, the amplitude of the oscillations with the phase can saturate to a finite value. This is in contrast to what we observe for optimized idle times $t_{\rm idle}=0.9/1$~ns, which decay to zero. To understand this feature, we carry out simulations of the dynamics of a qubit initialized in the $\frac{\ket{\downarrow}+i\ket{\uparrow}}{\sqrt{2}}$ superposition state which is shuttled between two neighboring quantum dots. Each shuttling step is modelled by a rotation. This rotation arises from the precession around the quantization axis of the quantum dot towards which the qubit is shuttled. We also calculate for every even $n$ the expected measurement result, i.e. the amplitude of the P$_{\uparrow}$ oscillations that appear when the phase $\upphi$ of the second $\pi/2$ pulse is varied. This is shown in \ref{figSup:effect_wrong_waittime}.c, with  two examples corresponding to a non-optimized idle time and an optimized idle time.

\ref{figSup:effect_wrong_waittime}.a displays the trajectory in the Bloch sphere of the qubit for the first 16 shuttling steps, in the reference frame of the quantum dot where the shuttling experiment starts. The different states of the qubit map a circle which is tilted compared to the equator. The product of the two rotations generated by shuttling back and forth is equivalent to a single rotation around a fixed axis. Consequently, multiple shuttling cycles can be seen as successive rotations around this fixed axis which elucidates the trajectory observed in the Bloch sphere. This also explains the oscillations of the amplitude as function of $n$ seen in \ref{figSup:vary_waitQ3}, as the distance between origin and the projection of the state on $xy$-plane can vary significantly depending on the number of shuttles for an non-optimized idle time. In contrast, when the idle times are well-optimized, the qubit states are on the equator of the Bloch sphere and no oscillations of the amplitude with $n$ can be observed.

Next, we include the effects of dephasing in the simulations, by assuming that the qubit frequencies fluctuate between repetitions of a given experiment with a fixed $n$. We observe that the state of the qubit is spread along a circle with a distribution that becomes more uniform as $n$ increases, meaning when the qubit experiences more dephasing. An example is shown in \ref{figSup:effect_wrong_waittime}.b for $n=98$, corresponding to the data shown in \ref{figSup:effect_wrong_waittime}.c. The center of the circle, which is equivalent to the statistical average of the qubit state when the qubit is completely dephased, is not on the equator on Bloch sphere. This explains the finite amplitude observed in the measurements at large $n$. Except for the revival of the amplitude observed for $t_{\rm idle}=2.1$~ns, these simulations capture most of the features observed in \ref{figSup:vary_waitQ3}.

\bibliography{bib_shuttling}

\clearpage